\documentclass{aa}
\usepackage{graphicx}

\newcommand{\coo}{C$^{18}$O}
\newcommand{\cco}{$^{13}$CO}

\newcommand{\hcccn}{HC$_{3}$N}

\begin{document}

\title{Distribution of HNCO 5$_{05}$-4$_{04}$ in Massive Star-forming Regions}

\author{Juan Li\inst{1, 2, 3} \and Junzhi Wang\inst{1, 2} \and Qiusheng Gu\inst{1, 2} \and Xingwu Zheng\inst{1, 2}}

\institute{School of Astronomy \& Space Science, Nanjing
University, 22 Hankou RD, Nanjing 210093, China \email{lijuan@shao.ac.cn} \and Key Laboratory of Modern Astronomy and Astrophysics
(Nanjing University), Ministry of Education, Nanjing 210093, China
\and Shanghai Astronomical Observatory, CAS, 80 Nandan Road, Shanghai 200030, China}

\date{Received 18 December 2012 / Accepted 6 May 2013 }

\abstract{}
{The goal of this paper is to study the spatial distribution of HNCO in massive star-forming
regions, and investigate its spatial association with infrared sources, as well as physical conditions in region of HNCO emission.}
{We have mapped nine massive star-forming regions in HNCO 5$_{05}$-4$_{04}$ with
the Purple Mountain Observatory 13.7m telescope. The \coo\ maps of these sources were obtained
simultaneously.} {The HNCO emission shows compact distribution, with
emission peak centred on water masers. Nearly all the HNCO clumps show signs of
embedded mid-infrared or far-infrared sources. The FWHM sizes of HNCO clumps are
significantly smaller than \coo\ clumps but rather similar to \hcccn\ clumps.
We also found good correlation between the integrated intensities, linewidths and
LSR velocities of HNCO and \hcccn\ emission, implying similar excitation mechanism
of these two species. As such, collisional excitation is likely to be the dominant
excitation mechanism for HNCO 5$_{05}-4_{04}$ emission in galactic massive star-forming regions.}{}


\maketitle

\section{Introduction}

Interstellar isocyanic acid (HNCO) was first detected in Sgr B2
molecular cloud complex, where it was found to be spatially extended
and relatively strong (Snyder \& Buhl 1972; Churchwell et al. 1986;
Lindqvist et al. 1995; Kuan \& Snyder 1996; Dahmen et al. 1997).
Since its discovery, HNCO has been detected in various
molecular clouds, including the dark cloud TMC-1 (e.g. Brown 1981;
Jackson et al. 1984), as well as hot cores
in massive star-forming regions (e.g. MacDonald et al. 1996;
Helmich \& van Dishoeck 1997). Jackson et al.
(1984) proposed that HNCO was a dense gas tracer due to
coincidence of HNCO emission with regions of high density ($n \geq
10^6$ cm$^{-3}$). Zinchenko et al. (2000) reported a detection rate
of 70\% in a survey of 81 molecular clouds. HNCO has also been detected in some
extragalactic sources (Nguyen-Q-Rieu et al. 1991; Meier \& Turner
2005; Mart\'{\i}n et al. 2009).

Within the Galactic center region, obvious different distribution of
HNCO and \coo, which is thought to be a tracer of the total H$_2$
column density, suggests possible different chemical properties of the
different molecular complexes in the center of our Galaxy (Dehmen et
al. 1997; Lindqvist et al. 1995; Mart\'{\i}n et al. 2008). Based on
the morphology of the emission and the HNCO abundance with respect
to H$_2$, several authors made the hypothesis that HNCO could be a
good tracer of interstellar shocks (e.g., Zinchenko et al. 2000;
Meier \& Turner 2005; Minh \& Irvine 2006). Mart\'{\i}n et al.
(2008, 2009) conducted a multitransition study of 13 molecular
clouds towards the Galactic center and concluded that the HNCO/CS
abundance ratio might provide an useful tool in distinguishing
between the influence of shocks and radiation activity in nuclear
regions of galaxies. Rodr\'{\i}guez-Fern\'{a}ndez et al. (2010) test
the hypothesis by observing a low-mass molecular outflow where the
chemistry is dominated by shocks. Their results indicate that shocks
can actually produce the HNCO abundance measured in galactic nuclei,
providing a solid basis to previous suggestions that the extended
HNCO in galactic nuclei could trace large scale shocks.

Chemical models have also been developed to investigate how HNCO
forms. Both gas-phase reaction (e.g. Turner et al. 1999) and formation routes on grain
surfaces (e.g. Hasegawa \& Herbst 1993; Garrod et al. 2008) have been used to model HNCO
abundance. Tideswell et al. (2010) found that HNCO is inefficiently formed when only gas-phase
formation pathways are considered in the chemical network, and
surface routes are needed to account for its abundance. Quan et al.
(2010) reproduced the abundances of HNCO and its isomers in cold and
warm sources using gas-grain simulation, which contains both
gas-phase and grain-surface syntheses.

Zinchenko et al. (2000) mapped three molecular clouds in HNCO transitions and found that
HNCO emission is compact and centrally peaked. Limited by angular resolutions, the size of HNCO clouds,
the relationships with infrared sources, as well as the dominant excitation mechanism of HNCO emission
remain unknown. From this consideration it is clear that high-sensitivity observations are needed to
better understand the physical condition and chemical properties of HNCO clouds.

In this paper we present large-scale mapping observations of HNCO
and \coo\ toward strong sources detected in
Zinchenko et al. (2000) with the Purple Mountain Observatory 13.7m (PMODLH 13.7m)
telescope. We first introduce the observations and data reduction
in \S\ \ref{observation}. In \S\ \ref{result}, we present the
observational results. In \S\ \ref{discussion}, we discuss implication of our observations on
the excitation mechanism and chemistry of HNCO, followed by a summary in \S\
\ref{summary}.

\section{Observations And Data Reductions}
\label{observation}

We performed mapping observations of HNCO $5_{05}-4_{04}$ (109.905 GHz)
and \coo\ 1-0 (109.782 GHz) lines simultaneously with the
PMODLH 13.7m located in Delingha, China in January, 2011. The main beam
size of about 55\arcsec\ and the pointing
accuracy is estimated to be better than 9\arcsec. A new cryogenically cooled 9-beam SIS
receiver (3$\times$3 with a separation of 174\arcsec\ between the centers of adjacent beams)
working in the 85-115 GHz band were employed. A fast Fourier
transform spectrometers (FFTS) of 16,384 channels with bandwidth of 1
GHz was used for each beam, supplying a velocity resolution of about 0.21 km
s$^{-1}$. Typical system temperature was around 150-300 K, depending
on the weather conditions. Observations were made in on-the-fly
mode with nine beams. The telescope is drifting in azimuth (with a rate of 20\arcsec\
s$^{-1}$) and stepping in elevation (with a scan step of 15\arcsec). Several
maps were made and later combined to lower rms noise levels. The
mapping size was 10\arcmin\ $\times$ 10\arcmin\ for most sources.
Mapping center of all our sources are listed in Table 1.

Most of the sources were selected from dense cores showing strong HNCO $5_{05}-4_{04}$ emission in
the surveys of Zinchenko et al. (2000). W44, S140 and DR21S are also included in our sample. W44 is
a molecular cloud interacting with a supernovae remnant (SNR), which provides promising environment
for production of HNCO. Both S140 and DR21S are massive star-forming regions with strong \hcccn\
emission (Li et al. 2012). Table 1 lists information of the sources
that have been mapped in this study. The distances were determined
from an extensive literature search. Trigonometric parallax
distances were used if available, otherwise photometric
distances or kinematic distances based on rotation curve of Fich,
Blitz \& Stark (1989) were used.

The data processing was conducted using Gildas package\footnote{\tt
http://www.iram.fr/IRAMFR/GILDAS.}. A least-square fit to baselines
in the spectra was carried out with the first order polynomial. The
baseline slopes were removed for all the sources. The individual
spectra were averaged and the resulting spectra were
Hanning-smoothed to improve the S/N ratio of the data. The line
parameters are obtained by Gaussian fitting. We express the results
in the unit of main beam brightness temperature (T$_{mb}$) assuming the main beam efficiencies of 0.5 (Chen et al. 2010, 2012).

To search for mid-infrared (MIR) emission of young stellar objects
(YSOs), we used the {\it Midcourse Space Experiment} ({\it MSX})
Galactic plane survey between 6 and 25 $\mu m$ at 18\arcsec\ spatial
resolution (Price et al. 2001). For S140, where no {\it MSX} data are
available, the {\it AKARI}/IRC source catalogue
(Murakami et al. 2007; Onaka et al. 2007) of {\it AKARI} all sky
survey is used to search for MIR emission (9 and 18 $\mu m$) with
spatial resolution of about 6\arcsec.
To search for far-infrared (FIR) emission of YSOs, we used {\it
AKARI}/FIS bright source catalogue (Kawada et al. 2007) centered at
65, 90, 140 and 160 $\mu m$, with spatial resolutions range from
37\arcsec\ to 61\arcsec.

\section{Observing Results}
\label{result}

Figure 1 presents the HNCO $5_{05}-4_{04}$ and \coo\ 1-0 spectra of S140 and W44. Both of them are newly detected in HNCO.
The derived line parameters of HNCO $5_{05}-4_{04}$ and \coo\ 1-0 at (0, 0) are
presented in Table 2 and 3, respectively, including main beam brightness
temperature ($T_{mb}$), integrated line intensity ($\int T_{mb}$d$\nu
$), LSR velocity and full-width half-maximum (FWHM) linewidth of HNCO
$5_{05}-4_{04}$ emission. All the parameters are determined from Gaussian fitting.
Our derived parameters are similar but not identical to results of Zinchenko et al. (2000)
because their main beam size is only two-thirds of our observations.

HNCO $5_{05}-4_{04}$ maps were obtained for nine sources. In Figure 2 and 3, we present
contour maps of HNCO $5_{05}-4_{04}$ (solid lines) and \coo\
(dashed lines) overlayed on MSX 21.3 $\mu$m image in linear scale. Except Orion KL,
other eight sources have been mapped with the same telescope and toward the same positions
with HNCO in \hcccn\ 10-9 transition (90.979 GHz) (Li et al. 2012). The \hcccn\ contour maps (dotted lines)
were also overlayed for comparison.
The MIR and FIR sources from {\it AKARI} catalogues are marked by asterisk ($\ast$) and star ($\star$), respectively. Red filled squares are used to mark positions of water masers. Compared with the widespread \coo\ distribution, HNCO emission shows rather compact distribution, with emissions well concentrating on water masers. The HNCO emission peaks offset from the \coo\ in Orion A, W51M and S158. Recently improved resolution of infrared observations allow us to better understand the
relationship between HNCO clumps and infrared sources. Nearly all
the HNCO clumps are spatially coincident with MIR or FIR sources,
which usually trace the warm dust emission of the cocoons of the OB
star central exciting sources. The morphology and spatial distribution of HNCO $5_{05}-4_{04}$ emission are similar to \hcccn\ 10-9 transition, which is a dense gas tracer (e.g. Li et al. 2012).
These results imply that HNCO emission is emitted from a small
volume of warmer and denser gas located nearer to the embedded objects
than the \coo\ emission.

The size of clouds is characterized using beam deconvolved angular
diameter and linear radius of a circle with same area as the half
peak intensity:
\begin{equation}
\theta_{transition} = 2(\frac{A_{1/2}}{\pi} -
\frac{\theta_{beam}^2}{4})^{1/2},
\end{equation}
\begin{equation}
R_{transition} = D(\frac{A_{1/2}}{\pi} -
\frac{\theta_{beam}^2}{4})^{1/2},
\end{equation}
where $A_{1/2}$ is the area within the contour of half peak
intensity, $\theta_{beam}$ is the FWHM beam size, and $D$ is the
distance of the source. The deconvolved linear FWHM sizes
and angular diameter of HNCO and \coo\ clouds are presented in Table 2 and 3, respectively. The deconvolved linear sizes of HNCO clouds range from 0.05 to 2.88 pc, with the
smallest value comes from Orion KL, and the largest value comes from
W51M. The sizes of HNCO clouds are significantly smaller than those
of \coo\ for nearly all the sources except W75OH, in
which the size of HNCO cloud is comparable to that of \coo.

Below we give comments on individual sources.

G121.30+0.66. We detected HNCO 5$_{05}$-4$_{04}$ emission
with T$_{mb}$ of 0.24 K. The HNCO clump has a centrally condensed
structure, with emission peak coincides with the MIR emission.

Orion KL. Orion KL is the closest massive star-forming region (414
pc; Menten et al. 2007). Several molecular components, such as the
Hot Core, Compact and Extended Ridges, and several luminous IR
sources or radio sources are associated with Orion KL. The nature of
the sources responsible for the luminous IR emission is still poorly
known and much debated. Other than being powered by an embedded
central heating source (e.g., Kaufman et al. 1998), an interesting
alternative explanation for this region's energetics is a
protostellar merger event that released a few times $10^{47}$ erg of
energy about 500 years ago (Bally \& Zinnecker 2005; Zapata et al.
2011; Bally et al. 2011). Our observations show that the HNCO peak
displace from the \coo\ peak and is centered on {\it MSX} 21.3 $\mu m$
emission peak. The HNCO/\coo\ intensity ratio at (0, 0) is
close to unity, while they are less than 0.25 in other sources.
As shock enhancement of HNCO has been detected in L1157 molecular
outflow (Rodr\'{\i}guez-Fern\'{a}ndez et al. 2010), the relatively high
HNCO/\coo\ intensity ratio seem to favor the idea that
the chemistry in Orion KL should be shock driven with much higher
temperature than the chemistry in a typical hot core (Favre et al.
2011), and produce high abundances of HNCO (Zinchenko et al. 2000).

W44. W44 is a SNR with shell-like morphologies
(Dubner et al. 2000; Jones et al. 1993). It locates adjacent to
giant molecular cloud with which it is suspected to be interacting
(Wootten 1981; Denoyer 1983). It contains OH 1720 MHz masers
attributed to dense shocked gas (Lockett et al. 1999). The {\it MSX}
image of W44 consists of a SNR and dust emission heated by embedded YSOs.
If it is correct that HNCO is enhanced in the presence of shocks due to its
injection in the gas phase from the grain mantles (Zincheko et al. 2000),
HNCO would be expected in molecular clouds interacting
with SNRs. Strong HNCO emission was detected with T$_{mb}$ of 0.58 K. The HNCO emission mainly concentrates around the YSO. More sources should be observed to investigate whether HNCO
is abundant in molecular clouds interacting with SNRs.

W51M. This is a strong HNCO 5$_{05}$-4$_{04}$ emission source with T$_{mb}$ of 0.48 K. Two cores were seen in HNCO clouds.
The western peak, W51m-west, is stronger than the eastern peak in HNCO emission,
but weaker than the eastern peak in \hcccn\ and \coo\ emission. W51m-west is
centered on the {\it AKARI} FIR sources, while the eastern peak is coincident with water
masers and near to MIR sources.

ON1. We detected HNCO 5$_{05}$-4$_{04}$ emission with T$_{mb}$ of 0.22 K.
Similar to G121.30+0.66, the HNCO emission morphology has a
centrally condensed structure, with the emission peak coincides
with the MIR emission.

DR21S. DR21S is a strong \hcccn\ emission source in Li et al. (2012). This is a
massive clump in the Cygnus X region. HNCO was detected in DR21S with T$_{mb}^*$ of
0.12 K. The HNCO emission peak is near to the MIR emission peak position.

W75OH. W75OH is also a massive clump in the Cygnus X region, which
hosts three massive dense cores at an early stage of their
evolution. We detected HNCO 5$_{05}$-4$_{04}$ emission
with T$_{mb}$ of 0.38 K. Two HNCO cores were seen in W75OH. The sourthern peak is
brighter than the northern peak in HNCO emission. The southern peak is near to the {\it AKARI}
FIR source, while the northern clump, W75OH-north, is near to the strong MIR emission.
Csengeri et al. (2011) reported detection of velocity
shears and low-velocity shocks in W75OH. Since HNCO is thought to be enhanced by
low-velocity shocks, high resolution observations of this source is expected to test
this hypothesis by comparing the spatial distribution of HNCO and N$_2$H$^+$ convergent flows.

S158. S158 is also referred as NGC7538 (Moreno \&
Chavarr\'{\i}a-K 1986). We detected HNCO 5$_{05}$-4$_{04}$ emission
with T$_{mb}$ of 0.44 K. Two clumps were seen in HNCO and \hcccn\ clouds, with the
stronger one coincident with the {\it AKARI} MIR source.

S140. S140 shows strong \hcccn\ emission in Li et al. (2012). HNCO 5$_{05}$-4$_{04}$ was marginally detected with T$_{mb}$ of 0.16 K. Two HNCO cores were seen with low signal-to-noise ratio. One of them is associated with the {\it AKARI} MIR sources, while another one is starless.

\section{Discussions}
\label{discussion}

Churchwell et al. (1985) mapped 14 transitions of HNCO toward Sgr B2. Analysis of population distribution indicates the FIR radiation from warm dust is likely to be responsible for the excitation of HNCO. They proposed that the most likely excitation mechanism of HNCO in Sgr B2 was radiative rather than collisional. In this case HNCO was a good probe of the FIR radiation field but not of gas properties such as density and kinetic temperature. The physical environment of hot cores differ from the Galactic center, in which the gas density (10$^3$-10$^4$ cm$^{-3}$) is much lower than in hot cores, thus the dominant excitation mechanism might be different. Zinchenko et al. (2000) explained the $K_{-1} > 0$ ladders of Orion KL with radiative excitation. For the $K_{-1} =0$ transitions with larger source size, Zinchenko et al. (2000) proposed that the radiative excitation would become inefficient, and the collisional excitation may dominate.

In Figure 4, we present plots comparing integrated intensities, linewidths, FWHM size and LSR velocities of HNCO 5$_{05}$-4$_{04}$ and \hcccn\ 10-9 (Li et al. 2012)\footnote{\tt We obtain T$_{mb}$ of \hcccn\ by assuming the main beam efficiencies of 0.5.}. The source names are also labeled in the figure. Good correlation is found for integrated intensities (r$_{corr}$ =0.63) and FWHM sizes (r$_{corr}$ =0.98) of HNCO and \hcccn\ clumps. The linewidths of HNCO clumps also agree well with \hcccn\ clumps (r$_{corr}$ =0.92), with the only exception of S140, which is possibly caused by the marginal detection of HNCO ($< 3\sigma$) in this source (see Figure 1). The velocity difference between HNCO and \hcccn\ are within 3$\sigma$ for all sources except W51M, which is possibly related to the complex structure of this source. The obvious correlations between line parameters as well as the similar morphology of HNCO and \hcccn\ clumps suggest that these two molecule lines are tracing a similar volume of gas. The critical density of \hcccn\ 10-9 is 10$^{6}$ cm$^{-3}$ (e.g., Chung et al. 1991), which is comparable with the critical density of HNCO 5$_{05}$-4$_{04}$. The dominant excitation mechanism for \hcccn\ is collisional excitation, thus collisional excitation is likely to be the dominant excitation mechanism for HNCO $K_{-1} =0$ emission in galactic massive star-forming regions.

In sources mapped by Zinchenko et al. (2000), HNCO emission peaks are significantly displaced from any known IR sources. However, observations at that time were severely limited by resolution of infrared observations. With greatly improved spatial resolution and sensitivity of IR observations, we found that nearly all the HNCO clumps in our observations are associated with MIR or FIR emission. Thus HNCO should be produced in hot gas. This is consistent with Bisschop et al. (2007), in which HNCO was classified as "hot" molecules based on rotation diagram analysis. Sanhueza et al. (2012) observed a sample of infrared dark clouds (IRDCs) and found that HNCO profiles show no evidence of being a tracer of shocks in most of the sources, only in two sources, which represent 10\% of the sources with HNCO detection, the HNCO spectrum presents a blue wing that is also observed in SiO. For results present here, we found possible shock enhancement of HNCO in two sources (Orion KL and W75OH). We conclude that HNCO is produced in warm environment, while shock could enhance the HNCO abundance.

\section{Summary and Prospects}
\label{summary}

In this paper we present HNCO $5_{05}-4_{04}$ mapping observations of nine massive
star-forming regions with the PMODLH 13.7m telescope. The \coo\ 1-0 maps of
these sources are obtained simultaneously. We used {\it MSX} and
{\it AKARI} satellite data to search for infrared emission from
YSOs and investigate the spatial relationship between HNCO clumps and infrared
sources.

We found good correlations between line parameters of HNCO and \hcccn\ emission, implying similar excitation mechanism of these two molecules. The size of HNCO clumps are much smaller than the \coo\ clumps and comparable with the \hcccn\ clumps. Thus collisional excitation is likely to be the dominant excitation mechanism for HNCO 5$_{05}$-4$_{04}$ transition in galactic massive star-forming regions.

We found that HNCO emission is compact and centrally condensed. Nearly all the HNCO clumps show signs of embedded infrared emission, supporting the idea that HNCO is a "hot" molecules. Future high resolution observations of W75OH is expected to test
the hypothesis that HNCO is enhanced by low-velocity shock by comparing the spatial distribution of HNCO and N$_2$H$^+$ convergent flows.


PMO is carrying out CO, \cco\ and \coo\ 1-0 survey toward the Galactic
plane with the DLH 13.7m telescope. HNCO 5$_{05}$-4$_{04}$ could be
observed simultaneously within the 1 GHz band, which would enable us to obtain spatial
distribution of strong HNCO emission in the Galactic plane and better study the chemical properties of HNCO.

\begin{acknowledgements} We thank the refree for his/her helpful comments and constructive suggestions. This work is partly supported by China Ministry of Science and Technology
under State Key Development Program for Basic Research (2012CB821800),
and partly supported by the Natural Science
Foundation of China under grants of 11103006, 10833006 and 10878010.
We would like to thank Key Laboratory of Radio Astronomy,
Chinese Academy of Sciences. We are very grateful to the staff of
Qinghai Station of Purple Mountain Observatory for their assistance
with the observations and data reductions. This research made use of
data products from the {\it Midcourse Space Experiment}. This
research is based on observations with {\it AKARI}, a JAXA project
with the participation of ESA. This research has made use of the
NASA/IPAC Infrared Science Archive, which is operated by the Jet
Propulsion Laboratory, California Institute of Technology, under
contract with the National Aeronautics and Space Administration.
\end{acknowledgements}



\clearpage

\begin{table}
\scriptsize
    \begin{center}
      \caption{Source List.}\label{tab:source}
      \begin{tabular}{lcccccc}
      \\
    \hline
    \hline
source name      & RA(J2000)     & DEC(J2000)       & D (kpc)   & Ref.    \\
            \hline
 G121.30+0.66   &  00:36:47.51  & 63:29:02.1       &   1.2  &  1   \\
 Orion KL       &  05:35:14.47   &  -05:22:27.56   &  0.414   &  2   \\
 W44            &   18:53:18.50 &  01:14:56.7      &  3.7  &  3   \\
 W51M           &   19:23:43.86 &  14:30:29.4      &  5.41  & 4  \\
 ON1            &   20:10:09.14 &  31:31:37.4      &  2.57   & 5   \\
 DR21S          &   20:39:00.80 &  42:19:29.8      &  1.5   &  6 \\
 W75OH          &   20:39:01.01 &  42:22:49.9      &  1.5   &  6  \\
 S158           &  23:13:44.84   &  61:26:50.71    &  2.65   &   7  \\
 S140           & 22:19:19.04   &  63:18:50.4       &  0.76  &   8  \\
             \hline
      \end{tabular}
  \end{center}
References: (1) Plume et al. 1992; (2) Menten et al. 2007; (3)
Solomon et al. 1987;; (4) Sato et al. 2010; (5) Rygl et al. 2010;
(6) Rygl et al. 2011; (7) Moscadelli et al. 2009; (8) Hirota et al.
2008.
\end{table}

\begin{table*}
    \begin{center}
\caption{Observational results of HNCO 5$_{05}$-4$_{04}$ Transition.}
  \begin{tabular}{l c c cccccc}
    \hline
    \hline
Source Name     & T$_{mb}$ & $\int$T$_{mb}$d$\nu$  &  V$_{LSR}$  &  FWHM   &  R$_{HNCO}$  &  $\theta_{HNCO}$   \\
                & (K)         & (K km s$^{-1}$) & (km s$^{-1}$)  &  (km s$^{-1}$)   &  (pc)   &  (arcsec)   \\
                \hline
G121.30+0.66    & 0.24(.04)  &  0.62(.08) & -17.09(.13)  &  2.36(.40)  & 0.17  & 28     \\
 Orion KL       & 0.96(.12)  &  7.10(.26) &  7.94(.11)   &  5.99(.34)  & 0.03  &  11     \\
 W44            & 0.58(.08)  &  3.24(.22) &  58.27(.16)  &  5.20(.46)   &  0.67  & 37   \\
 W51M           & 0.48(.06)  &  4.80(.22) &  55.75(.22)  &  9.52(.51)   & 1.44 & 55      \\
 W51M-west      & 0.46(.06)  &  5.46(.22) &  64.58(.23)  &  11.28(.48)  &  -    &  -        \\
 ON1            & 0.22(.06)  &  1.12(.12) &  11.79(.26)  &  4.63(.59)  & 0.45  &  36    \\
 DR21S          & 0.12(.03)  &  0.46(.06) &  -2.17(.21)  &  3.50(.56)   & 0.29  & 39      \\
 W75OH-north    & 0.26(.04)  &  1.04(.10) &  -3.25(.17)  &  3.75(.46)   &  -    &   -    \\
 W75OH          & 0.38(.04)  &  1.60(.10) &  -3.31(.13)  &  4.02(.32)    & 0.50  &  68    \\
 S158           & 0.44(.06)  &  1.96(.12) &  -56.15(.13) &  4.23(.29)    & 0.69  &  53   \\
 S140           & 0.16(.06)  &  1.08(.16) & -7.10(.47)   &  6.03(1.21)    & 0.15 &  41    \\
\hline
  \end{tabular}
  \end{center}
  \label{tab:hnco}
\end{table*}

\begin{table*}
   \begin{center}
\caption{Observational results of \coo\ 1-0 Transition}
  \begin{tabular}{lcccccccc}
    \hline
    \hline
Source Name     & T$_{mb}$ & $\int$T$_{mb}$d$\nu$  &  V$_{LSR}$  &  FWHM   &  R$_{HNCO}$  &  $\theta_{HNCO}$   \\
                & (K)        & (K km s$^{-1}$) & (km s$^{-1}$)  &  (km s$^{-1}$)   &  (pc)   &  (arcsec)   \\
                \hline
G121.30+0.66    &   1.52(.06)    &  3.32(.06)   &  -17.33(.02)   &   2.04(.05)   &  0.43    &  73      \\
 Orion KL       &   1.76(.08)    &  7.72(.14)   &   8.84(.04)    &   4.13(.10)  &  0.16    &  78   \\
 W44            &   4.42(.08)    &  26.66(.26)  &   58.19(.03)   &   5.68(.07)    &  1.40    &    78 \\
 W51M           &  2.70(.08)    &  31.94(.06)  &   56.49(.06)   &  11.10(.11)  & 2.20   &   84     \\
 W51M-west      &  1.21(.08)    &  25.34(0.40) &   60.90(.16)   &   19.66(.34) & -    &    -     \\
 ON1            &  1.78(.06)    &  6.70(.12)   &  11.35(.03)    &   3.53(.07)   &  0.74    &   59  \\
 DR21S          &  2.92(.06)    &  9.60(.10)   &  -2.37(.01)    &   3.09(.04)   &  0.52    &   71    \\
 W75OH-north    &  3.22(.04)    &  9.66(.08)   &  -3.69(.01)    &   2.82(.03)   &  -       &   -    \\
 W75OH          &  2.06(.04)    &  11.22(.12)   &  -2.95(.02)    &   3.45(.04)   &  0.63    &   87    \\
 S158           &   1.38(.10)    &  7.98(.20)   &   -56.18(.07)  &   5.43(.17)   &  1.27    &  98 \\
 S140           &   2.08(.06)    &  6.54(.14)   &  -7.23(.03)    &   2.95(.07)   &  0.40    &  108 \\
\hline
  \end{tabular}
  \end{center}
  \label{tab:c18o}
\end{table*}

\begin{figure*}
\begin{center}
\includegraphics[width=2.2in]{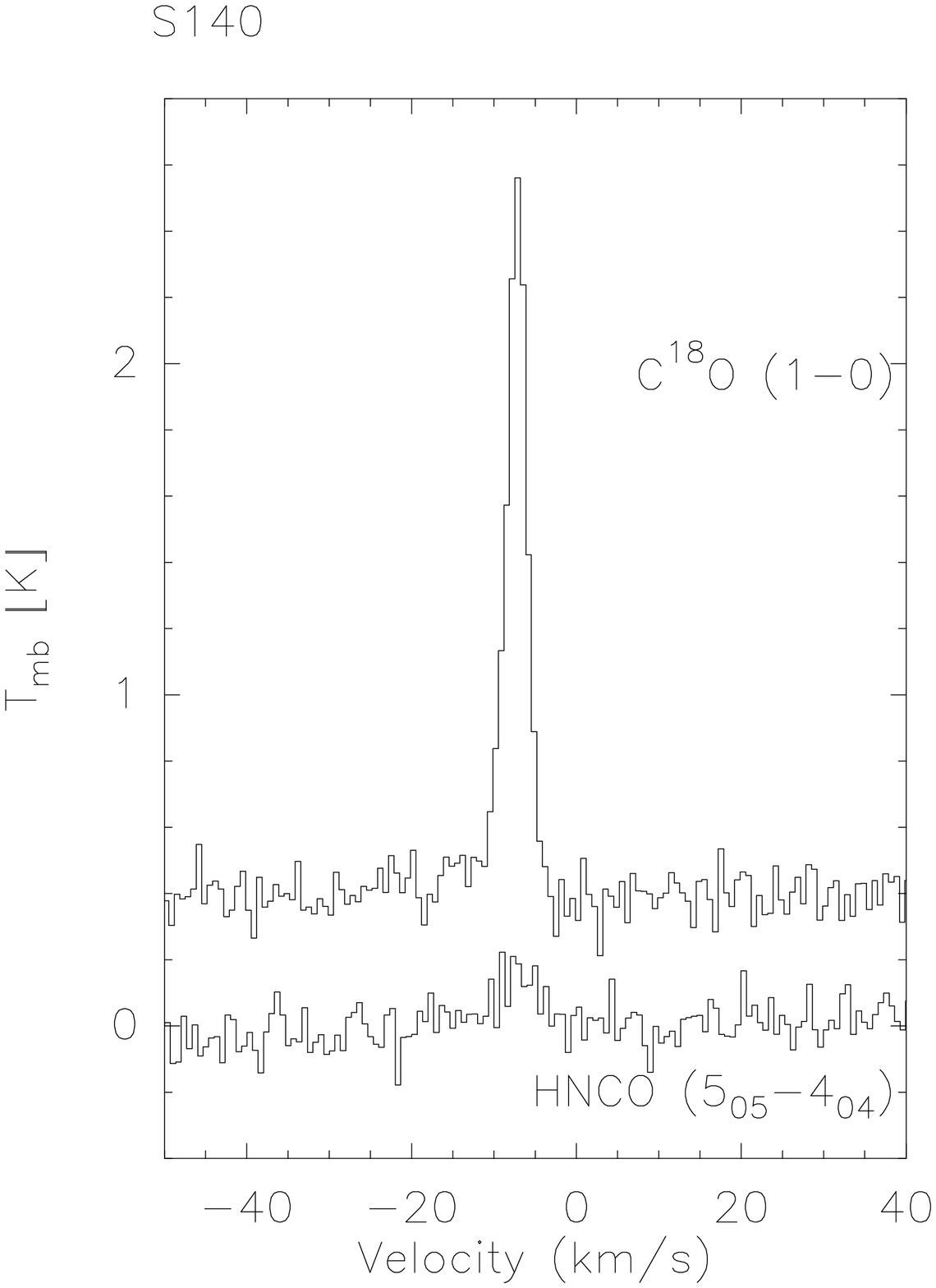}
\includegraphics[width=2.2in]{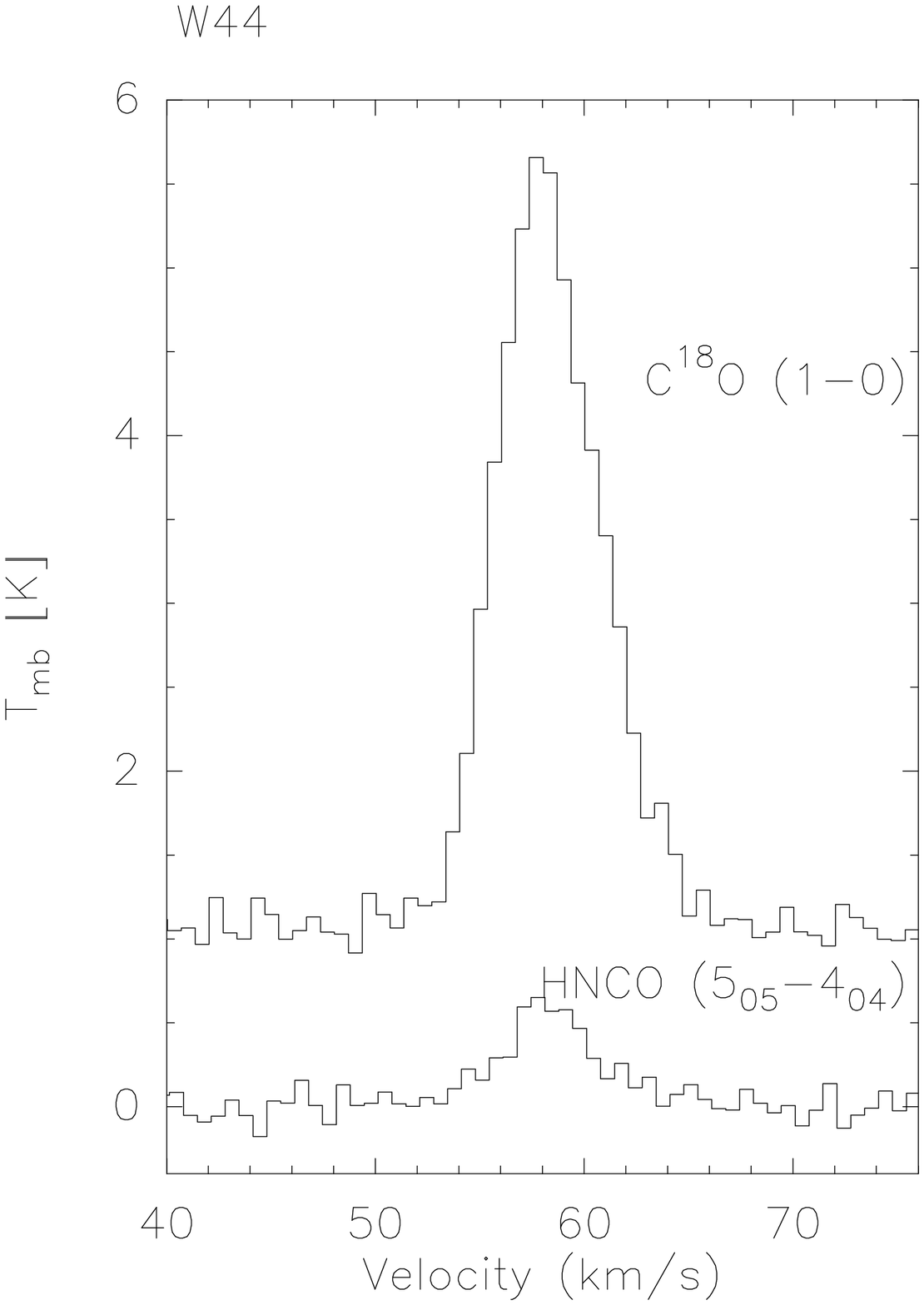}
\vspace*{-0.2 cm} \caption{\label{spec3} HNCO 5$_{05}$-4$_{04}$ and \coo\ 1-0 spectra of S140 and W44 at (0, 0). The identification of the transitions is given to the right of each lines.}
\end{center}
\end{figure*}

\begin{figure*}
   \centering
\includegraphics[scale=.32,angle=-90]{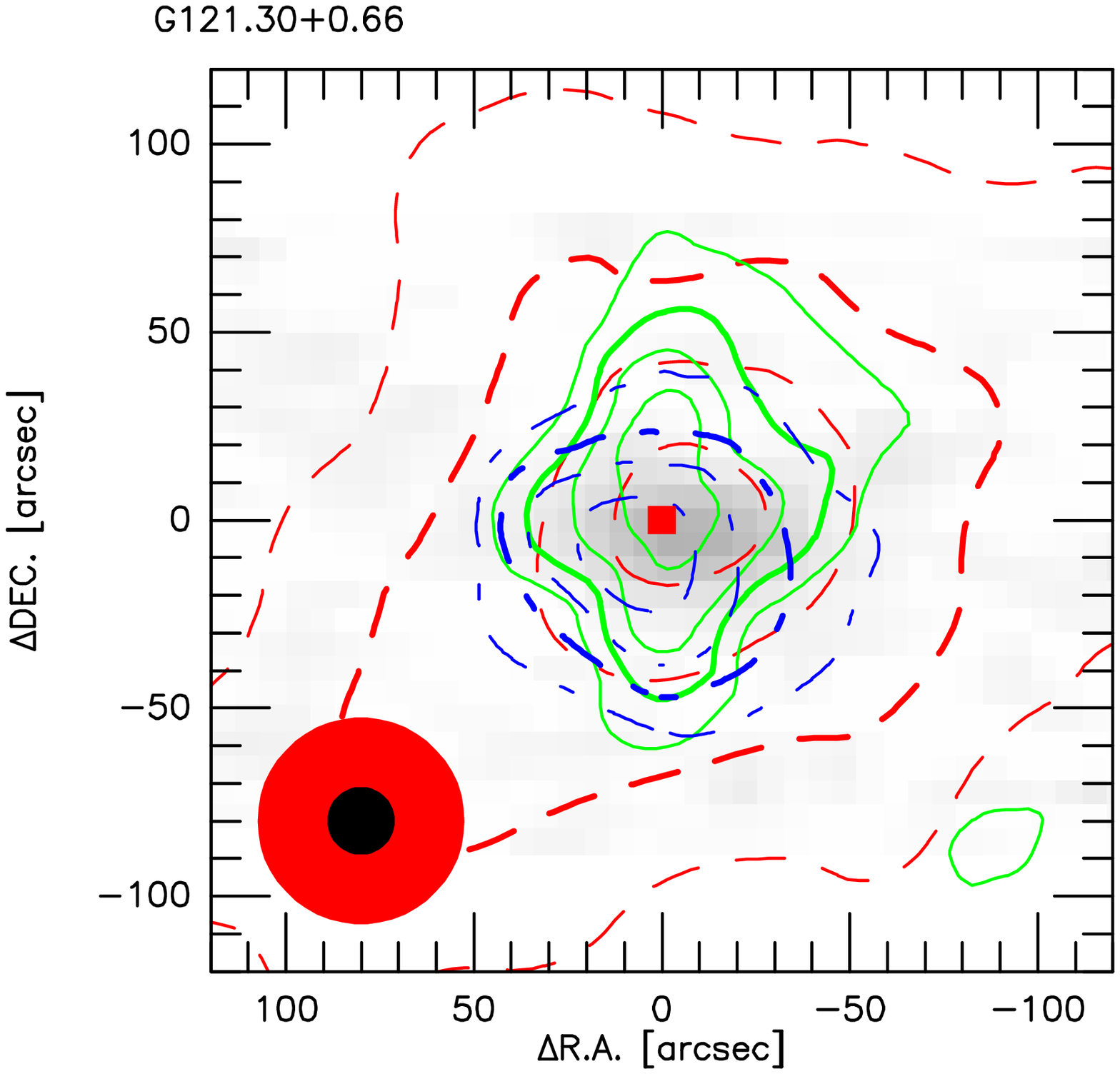}
\includegraphics[scale=.32,angle=-90]{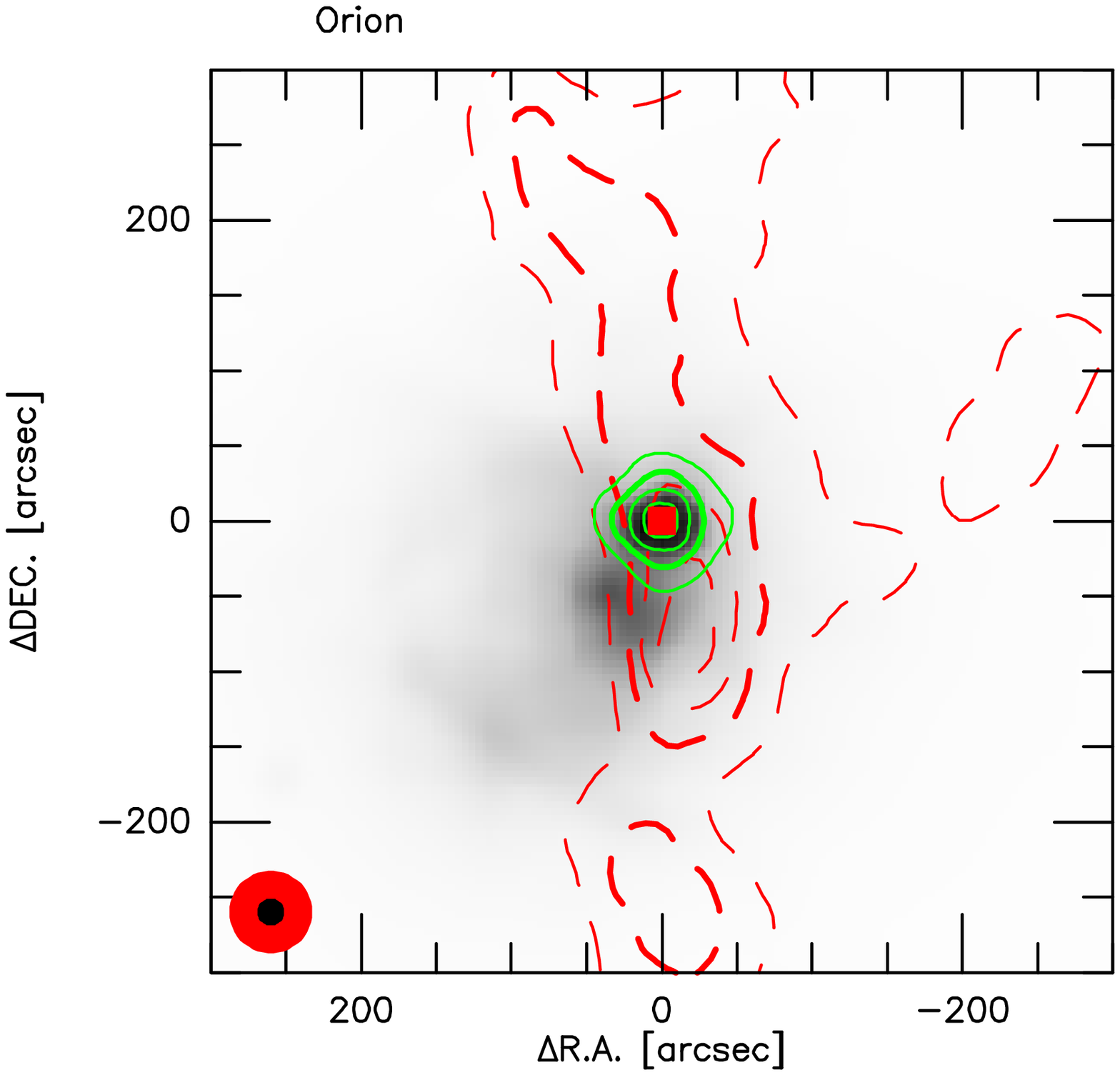}
\includegraphics[scale=.32,angle=-90]{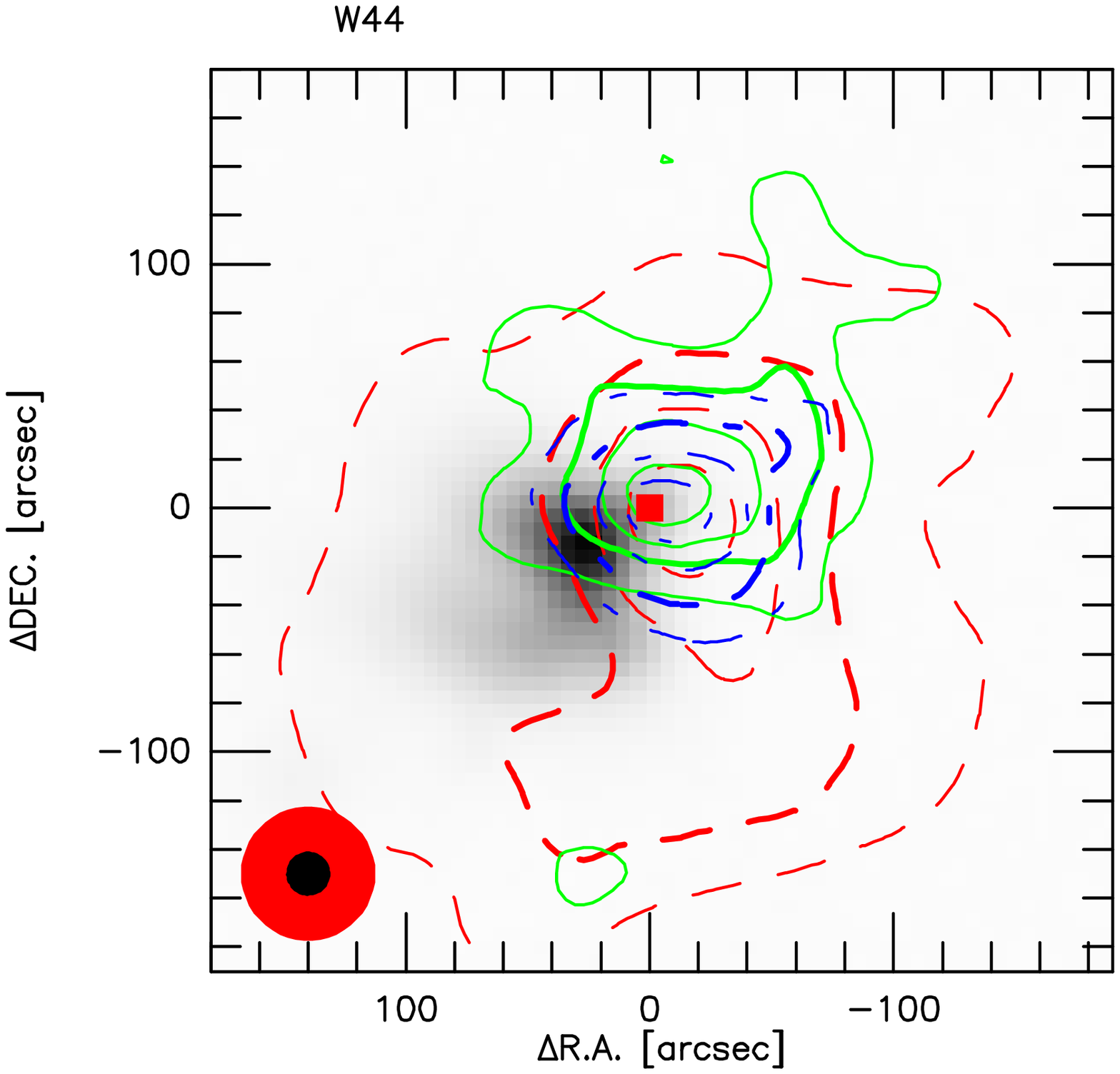}
\includegraphics[scale=.32,angle=-90]{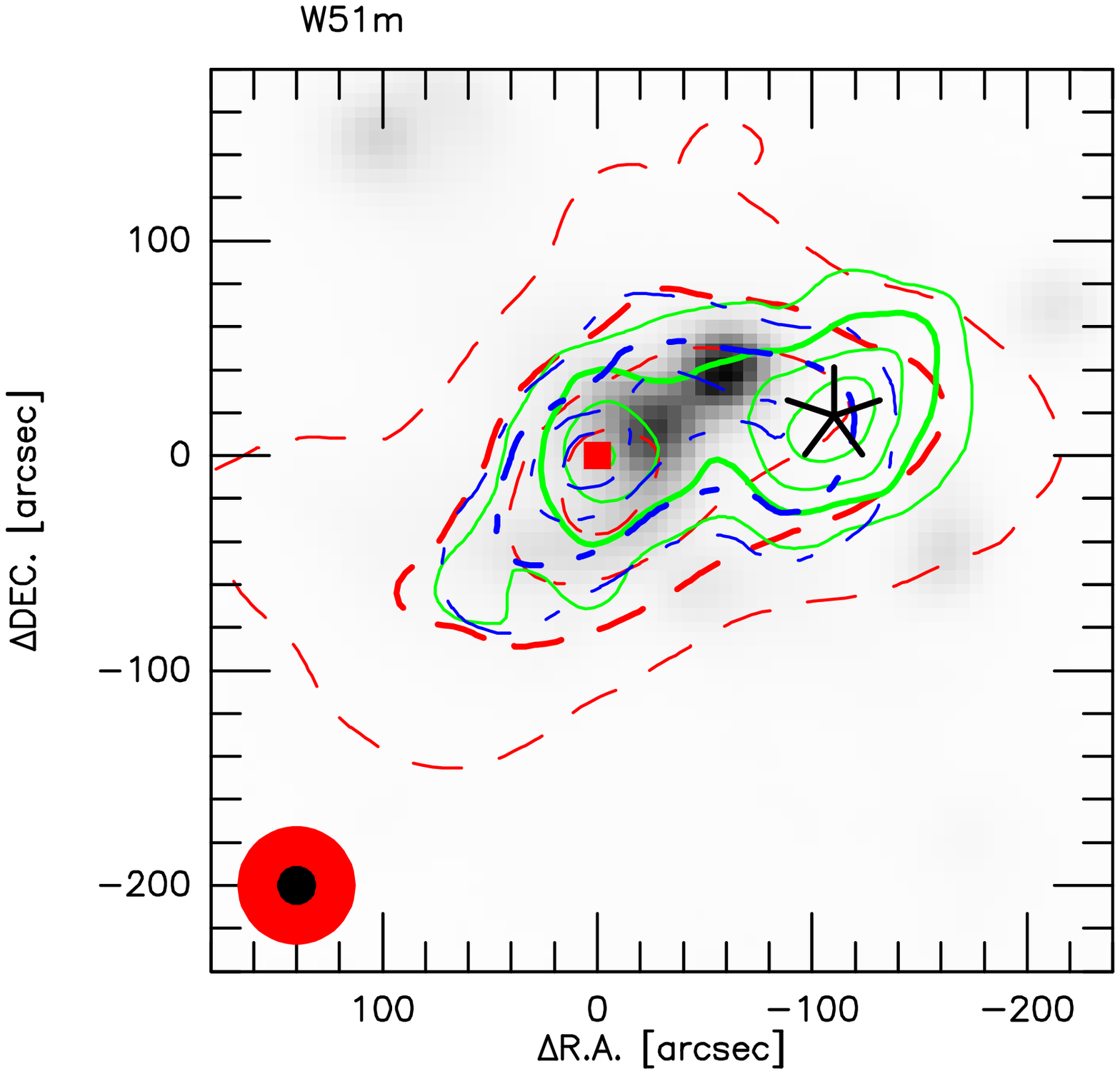}
\includegraphics[scale=.32,angle=-90]{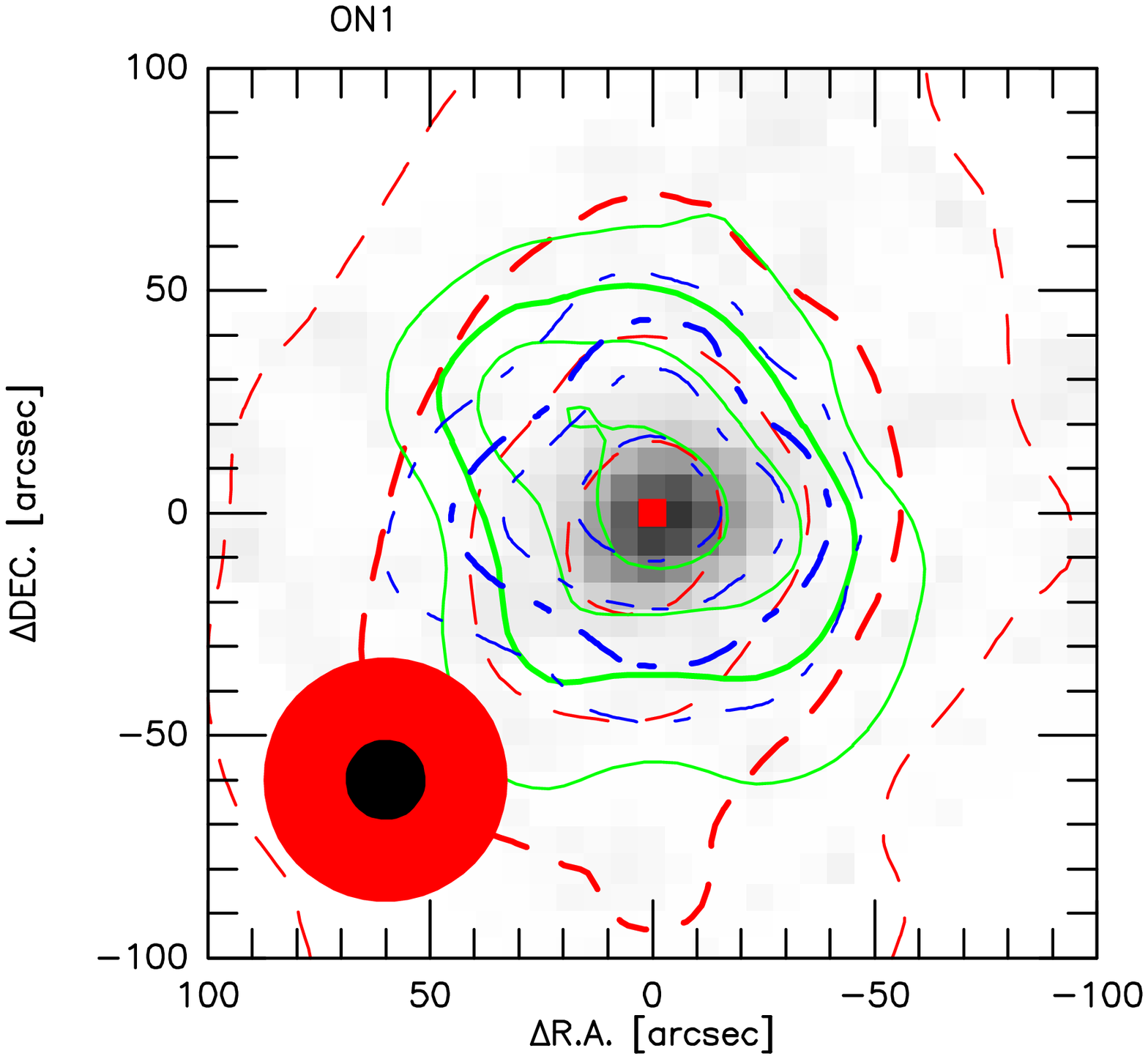}
\includegraphics[scale=.32,angle=-90]{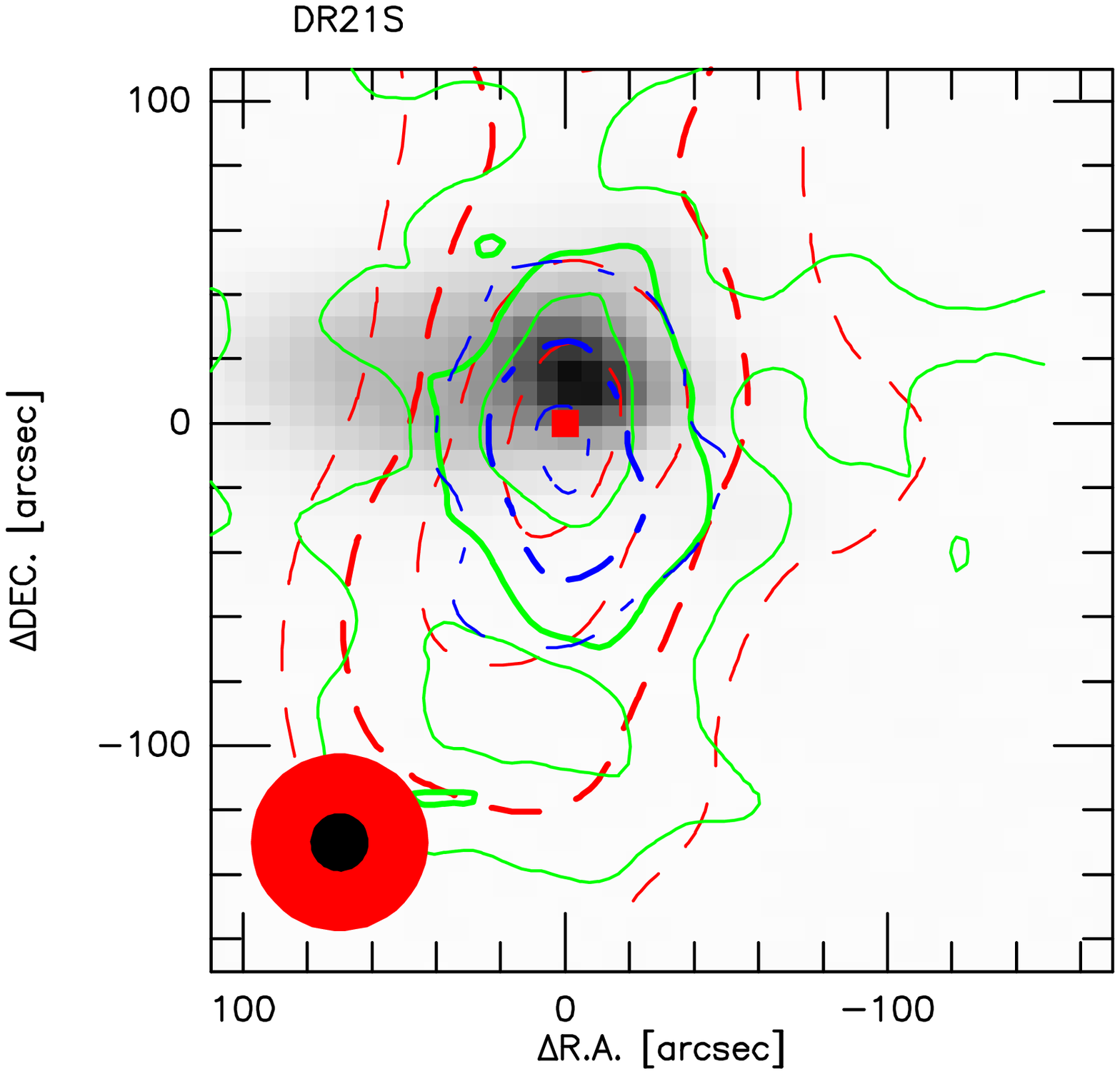}
\caption{Contour maps of HNCO (green solid line), \hcccn\ (blue dotted line) and \coo\ (red
dashed line) superimposed on {\it MSX} 21.3 $\mu$m image for
observing sources. The contour levels are 30\%, 50\%, 70\% and 90\% of the map peak, reported in Table 3 (see Column 3 for HNCO, Column 7 for \coo, see Column 3 in Table 3 of Li et al. (2012) for \hcccn). The heavy lines represent 50\% of the map peak. "$\star$"
is used to mark the position of the {\it AKARI} FIR source. "$\ast$" is used to mark the position of the {\it AKARI} MIR source in S140. Red filled squares are used to mark the position of water masers. The FWHM beam size for molecular lines (the big, red circle) and mid-infrared (the small, black circle) observations are shown at the lower left the maps. (A color version of this figure is available in the online journal.)}
\end{figure*}

\begin{figure*}
   \centering
\includegraphics[scale=.32,angle=-90]{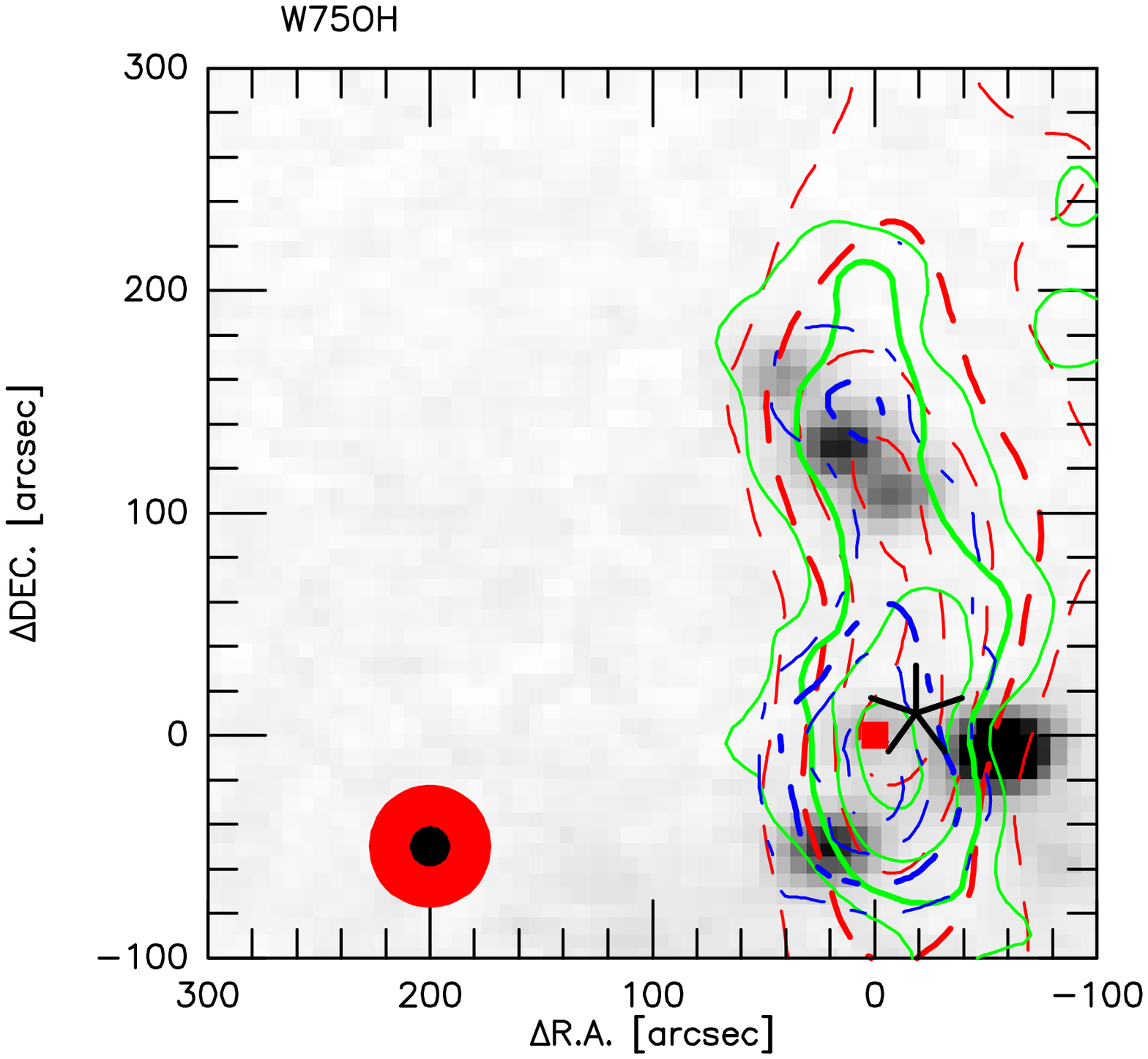}
\includegraphics[scale=.32,angle=-90]{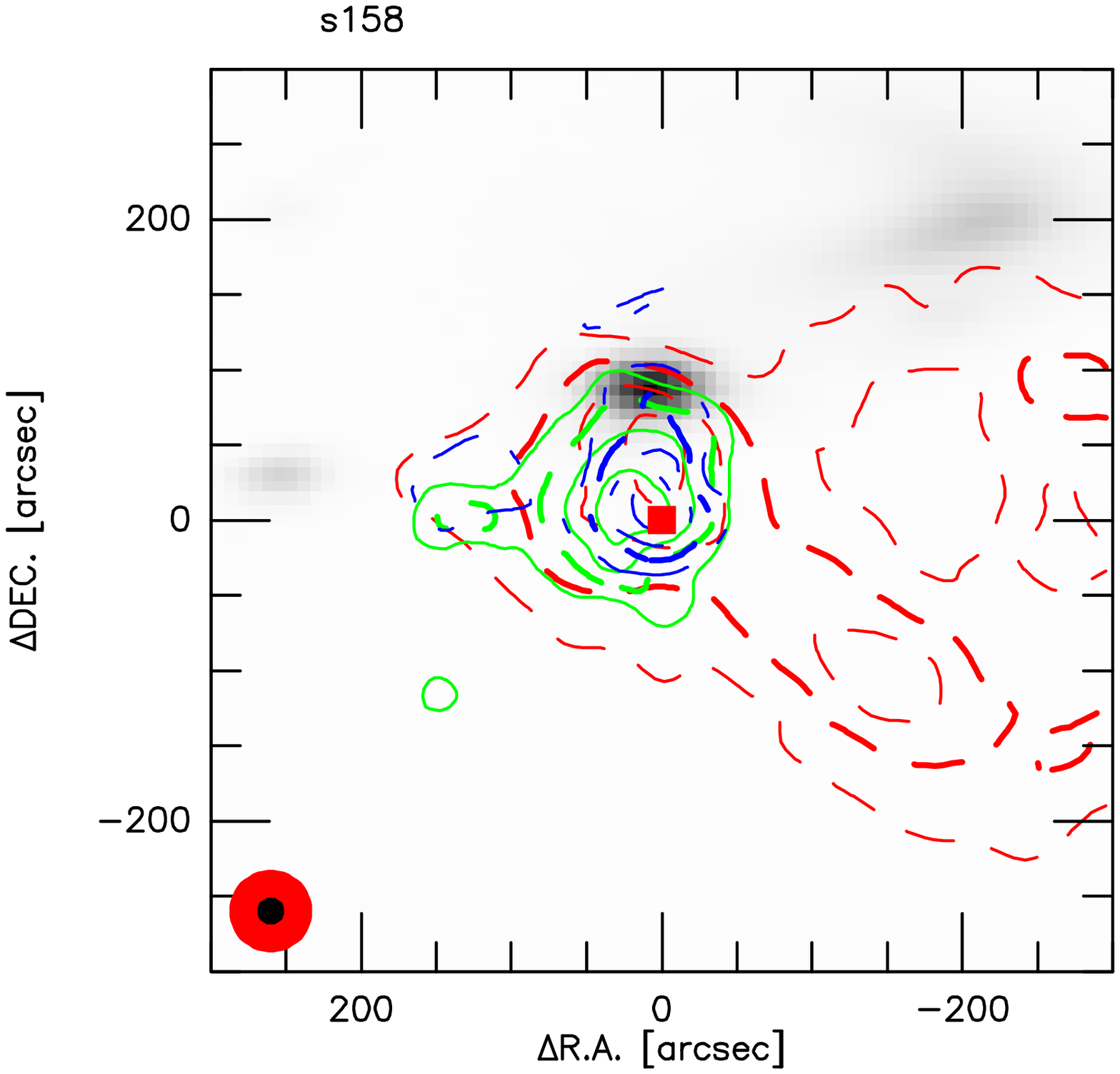}
\includegraphics[scale=.32,angle=-90]{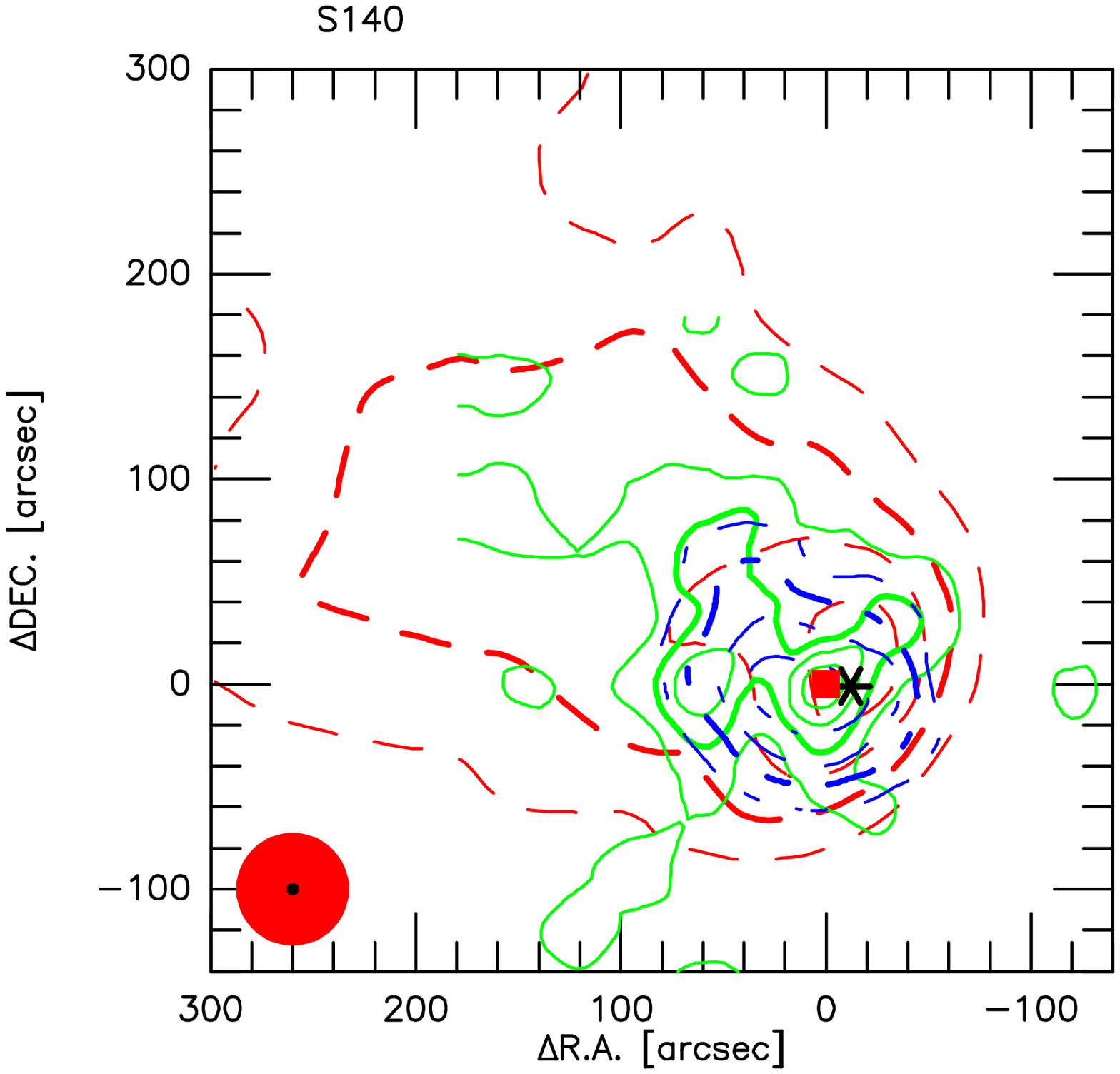}
\caption{Figure 2 continued. (A color version of this figure is available in the online journal.)}
\end{figure*}

\begin{figure*}
   \centering
\includegraphics[scale=.6]{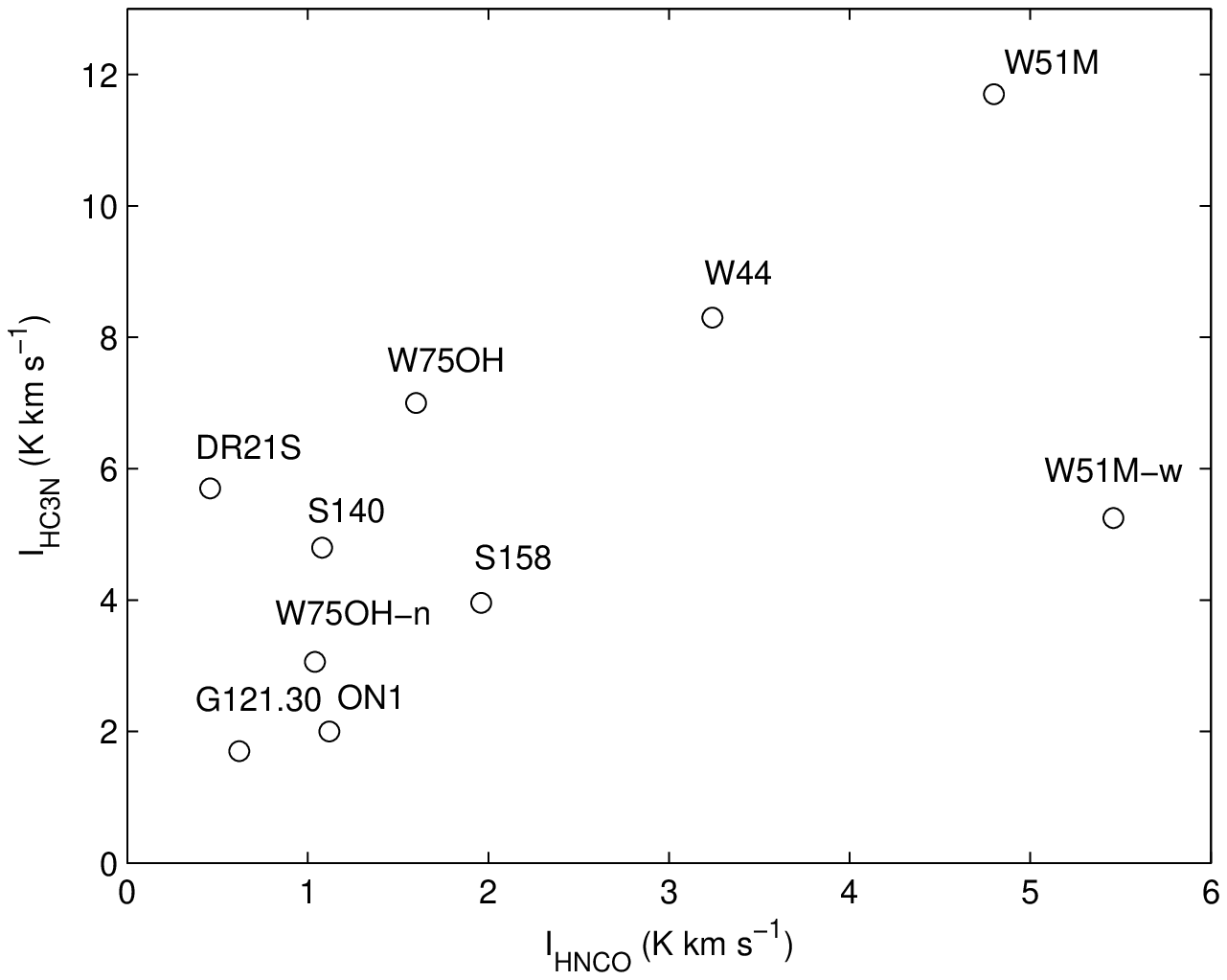}
\includegraphics[scale=.6]{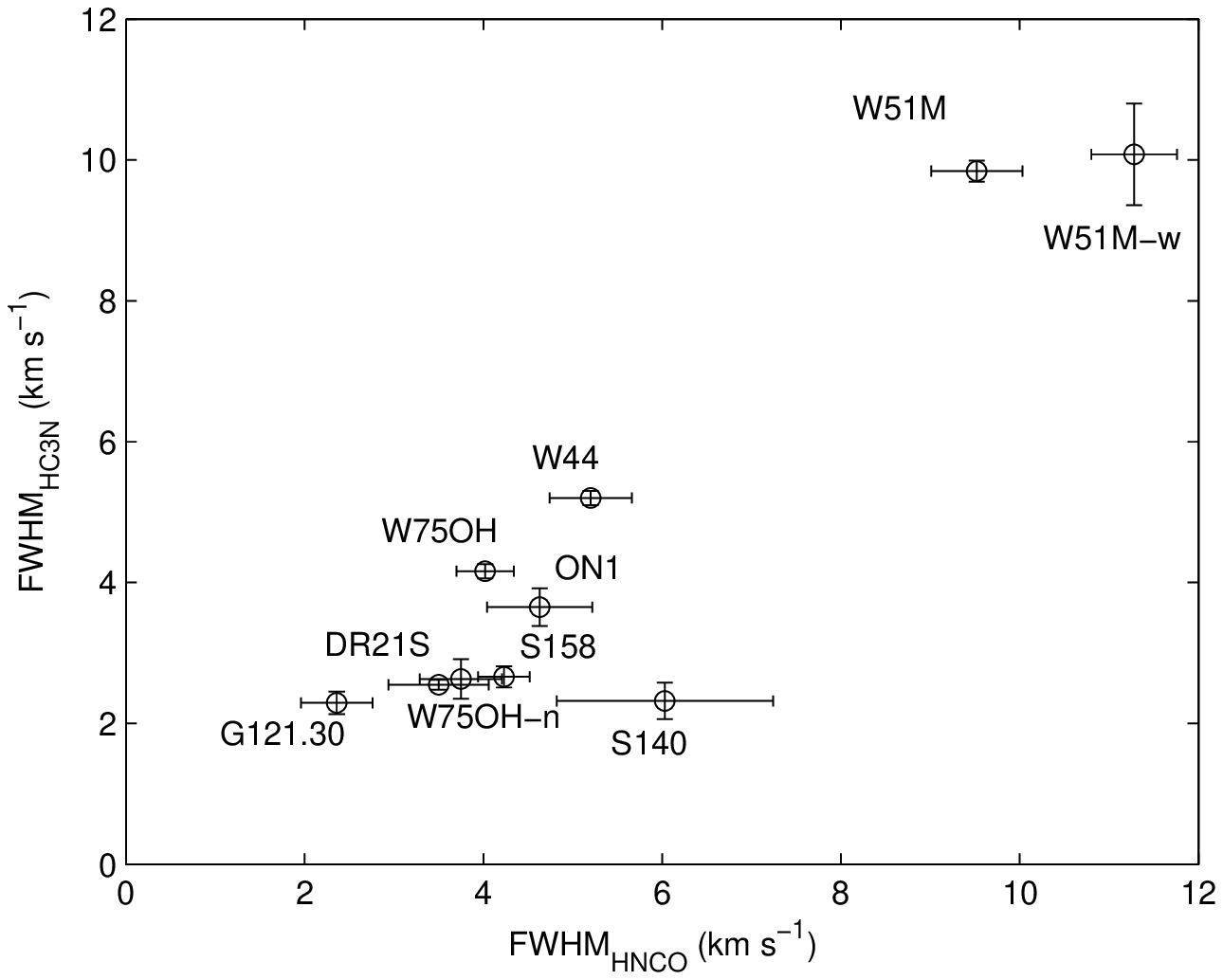}
\includegraphics[scale=0.6]{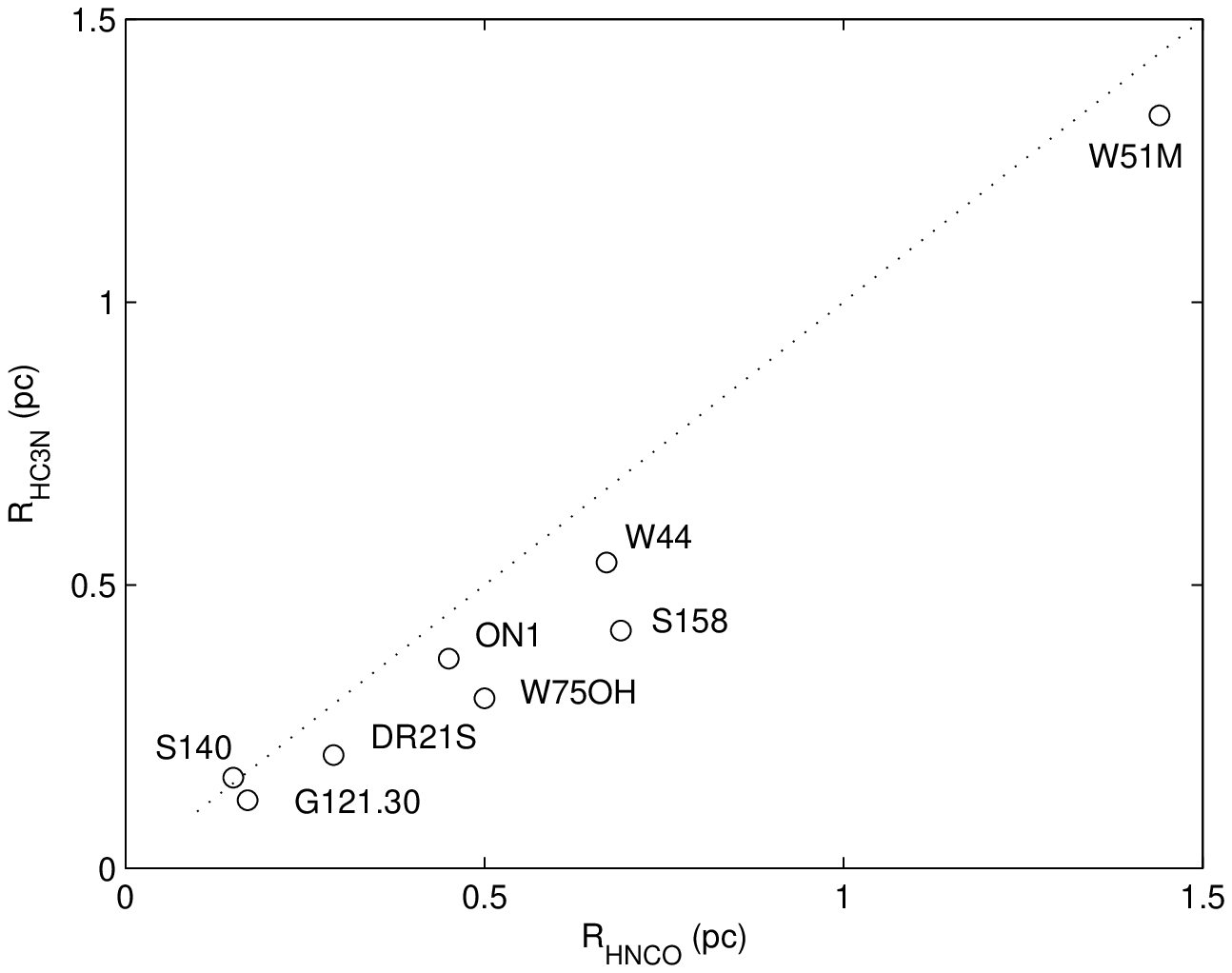}
\includegraphics[scale=.6]{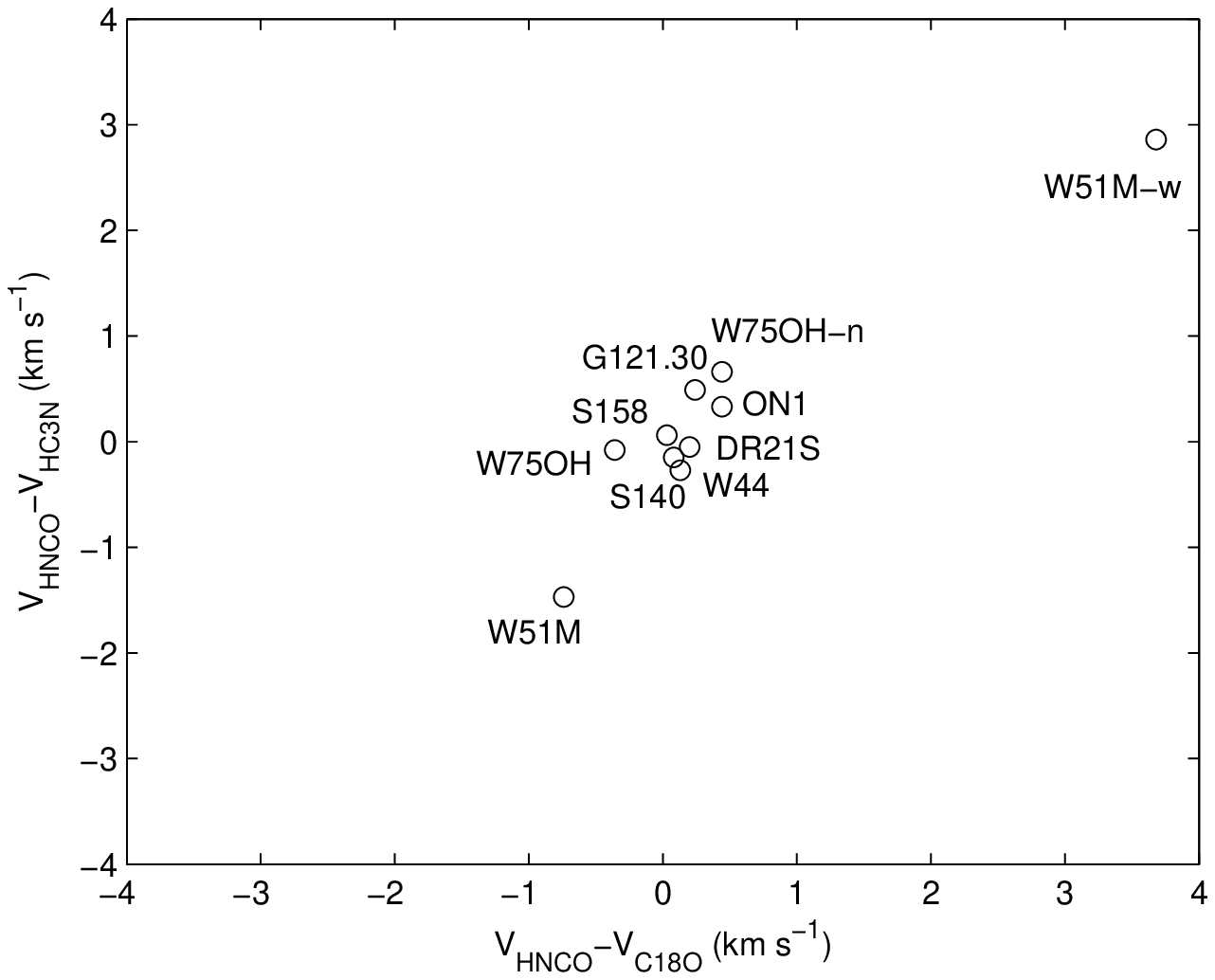}
\caption{ Upper left: comparison of integrated intensities of HNCO $5_{05}-4_{04}$ and \hcccn\ 10-9; upper right: comparison of linewidth of HNCO $5_{05}-4_{04}$ and \hcccn\ 10-9; lower left: comparison of size of HNCO $5_{05}-4_{04}$ and \hcccn\ 10-9. The dotted line has a slop of 1; lower right: LSR velocity difference between HNCO $5_{05}-4_{04}$ and \hcccn\ 10-9 versus LSR velocity difference between HNCO $5_{05}-4_{04}$ and \coo\ 1-0. The source names were also labeled in the figure. }
\end{figure*}


\begin{thebibliography}{99}

\bibitem[Bally et al. (2011)]{bally11}
Bally, J. Cunningham, N. J., Moeckel, N. et al. 2011, ApJ, 727, 113

\bibitem[Bally (2005)]{bally05}
Bally, J. \& Zinnecker, H. 2005, AJ, 129, 2281

\bibitem[Bisschop (2007)]{bisschop07}
Bisschop, S. E., Jorgensen, J. K., van Dishoeck, E. F. \& de Wachter, E. B. M. 2007, A\&A, 465, 913

\bibitem[Brown et al. (1981)]{brown81}
Brown, R. L. 1981, ApJ, 248, 119

\bibitem[Chen et al. (2010)]{chen10}
Chen, X., Shen, Z.-Q., Li, J. J. et al. 2010, ApJ, 710, 150

\bibitem[Chen et al. (2012)]{chen12}
Chen, X., Ellingsen, S. P., He, J.-H. et al. 2012, ApJS, 200, 5

\bibitem[Chung et al. (1991)]{chung91}
Chung, H. S., Kameya, O. \& Morimoto, M. 1991, JKAS, 24, 217

\bibitem[Churchwell et al. (1986)]{churchwell86}
Churchwell, E., Wood, D., Myers, P. C. \& Myers, R. V., 1986, ApJ,
305, 405

\bibitem[Csengeri (2011)]{csengeri11}
Csengeri, T., Bontemps, S., Schneider, N. et al. 2011, ApJ, 740, 5

\bibitem[Dahmen (1997)]{dahmen97}
Dahmen, G. et al. 1997, A\&AS, 126, 197

\bibitem[Denoyer (1983)]{denoyer83}
Denoyer, L. K. 1983, ApJ, 264, 141

\bibitem[Dubner (2000)]{dubner00}
Dubner, G. M., Vela\'{a}zquez, P. F., Goss, W. M. \& Holdaway, M. A. 2000, AJ, 120, 1933

\bibitem[Favre (2011)]{favre11}
Favre, C., Despois, D., Brouillet, N. et al. 2011, A\&A, 532, 32

\bibitem[Fich (1989)]{Fich89}
Fich, M. Blitz, L., Stark, A. A. 1989, ApJ, 342, 272

\bibitem[Garrod (2008)]{garrod08}
Garrod, R. T., Weaver, S. L. W. \& Herbst, E. 2008, ApJ, 682, 283

\bibitem[Hasegawa (1993)]{hasegawa93}
Hasegawa, T. I. \& Herbst, E. 1993, MNRAS, 261, 83

\bibitem[Helmich (1997)]{helmich97}
Helmich, F. P. \& van Dishoeck, E. F. 1997, A\&AS, 124, 205

\bibitem[Jackson (1984)]{jackson84}
Jackson, J. M., Armstrong, J. T. \& Barett, A. H. 1984, ApJ, 280, 608

\bibitem[Jones (1993)]{jones93}
Jones, L. R. Smith, A. \& Angellini, L. 1993, MNRAS, 265, 631

\bibitem[Kawada (2007)]{kawada07}
Kawada, M., Baba, H., Barthel, P. D. et al. 2007, PASJ, 59, 389

\bibitem[Kaufman (1998)]{kauffman98}
Kaufman, M. J., Hollenbach, D. J. \& Tielens, A. G. G. M. 1998, ApJ, 497, 276


\bibitem[Kuan et al. (1996)]{kuan96}
Kuan, Y. J., Snyder, L. E. 1996, ApJ, 470, 981

\bibitem[Li (2012)]{li12}
Li, J., Wang, J. Z., Gu, Q. S., Zhang, Z.-Y. \& Zheng, X. W. 2012, ApJ, 745, 47

\bibitem[Lindqvist (1995)]{lindqvist95}
Lindqvist, M., Sandqvist, A., Winnberg, A., Johansson, L. E. B. \&
Nyman, L.-A. 1995, A\&AS, 113, 257

\bibitem[Lockett (1999)]{lockett99}
Lockett, P., Gauthier, E. \& Elitzur, M. 1999, ApJ, 511, 235

\bibitem[MacDolnald (1996)]{macdonald96}
MacDolnald, G. H., Gibb, A. G., Habing, R. J. \& Millar, T. J. 1996, A\&AS, 119, 333

\bibitem[Martin et al. (2008)]{martin08}
Martin, S., Requena-Torres, M. A., Martin-Pintado, J. \&
Mauersberger, R. 2008, ApJ, 678, 245

\bibitem[Martin et al. (2009)]{martin09}
Martin, S., Martin-Pintado, J. \& Mauersberger, R. 2009, ApJ, 694,
610

\bibitem[Meier et al. (2005)]{Meier05}
Meier, D. S., \& Turner, J. L., 2005, ApJ, 618, 259

\bibitem[Menten et al. (2007)]{Menten07}
Menten, K. M., Reid, M. J., Forbrich, J. \& Brunthaler, A. 2007,
A\&A, 474, 515

\bibitem[Minh (2006)]{minh06}
Minh, Y. C. \& Irvine, W. M. 2006, NewA, 11, 59

\bibitem[Moreno (1986)]{morenno86}
Moreno, M. A. \& Chavarr\'{\i}a-K, C. 1986, A\&A, 161, 130

\bibitem[Murakami et al. (2007)]{Murakami07}
Murakami, H. et al. 2007, PASJ, 59, 369

\bibitem[Nguyen-q-rieu et al. (1991)]{Nguyen-q-rieu91}
Nguyen-Q-Rieu, Henkel, C., Jackson, J. M., Mauersberger, R. 1991, A\&A, 241, 33

\bibitem[Onaka et al. (2007)]{onaka07}
Onaka, T., Matsuhara, H., Wada, T. et al. 2007, PASJ, 59, 401

\bibitem[Price et al. (2001)]{price01}
Price, S. D., Egan, M. P., Carey, S. J. et al. 2001, AJ, 121, 2819

\bibitem[Quan et al. (2007)]{quan07}
Quan, D. H., Herbst, E., Osamura, Y. \& Roueff, E. 2010, ApJ, 725, 2101

\bibitem[Rodriguez et al. (2010)]{rodriguez10}
Rodr\'{\i}guez-Fern\'{a}ndez, N. J., Tafalla, M., Gueth, F. \& Bachiller, R.,
2010, A\&A, 516, 98

\bibitem[Sanhueza et al. (2012)]{sanhueza12}
Sanhueza, P., Jackson, J. M., Foster, J. B. et al. 2012, ApJ, 756, 60

\bibitem[Snyder et al. (1972)]{snyder72}
Snyder, L. E. \& Buhl, D. 1972, ApJ, 177, 619

\bibitem[Tideswell (2010)]{tideswell10}
Tideswell, D. M., Fuller, G. A., Millar, T. J. \& Markwick, A. J. 2010, A\&A, 510, 85

\bibitem[Turner (1999)]{turner99}
Turner, B. E., Terzieva, R. \& Herbst, E. 1999, ApJ, 518, 699

\bibitem[Wootten (1981)]{wootten81}
Wootten, A. 1981, ApJ, 245, 105

\bibitem[Zapata et al. (2011)]{zapata11}
Zapata, L. A., Schmid-Burgk, J. \& Menten, K. M. 2011, A\&A, 529, 24

\bibitem[Zinchenko et al. (2000)]{zinchenko00}
Zinchenko, I., Henkel, C. \& Mao, R. Q. 2000, A\&A, 361, 1079

\end{thebibliography}
\end{document}